\documentclass[aps,prl,twocolumn]{revtex4} 
\usepackage{graphicx}% Include figure files

\begin{document}
\title{Universal level-spacing distribution in quantum systems}
\author{Viktor A. Podolskiy} 
\author{Evgenii E. Narimanov}
\address{Electrical Engineering Department, Princeton University, Princeton, NJ 08544, USA}

\maketitle

\begin{figure}
\includegraphics[width=6cm]{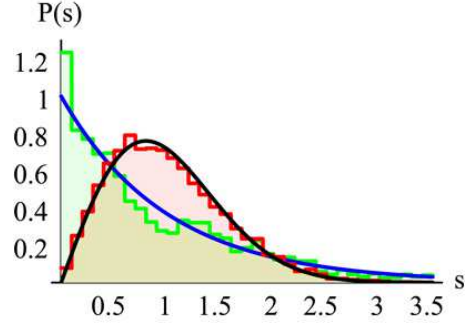}
\caption{\label{figdist}
Level spacing distribution in integrable system (green histogram and blue curve) is Poissonian, while it changes to Wigner when the system becomes completely chaotic (red, black). We compare the exact analytical formulae [Eqs.(\ref{eqPoisson}),(\ref{eqWigner})] (blue, black) to numerically simulated energy levels in the optical microcavities (green, red).
}
\end{figure}

{\bf 
Classical counterparts of a great variety of quantum systems, from atomic physics \cite{SpacingHe,DTNature,DTScience} to quantum wells \cite{Wells} and quantum dots\cite{Dots}, to optical\cite{BTScience,WGNature}, microwave\cite{KodroliSridhar1994,Alt1994}, and acoustic resonators\cite{SpacingAl} exhibit partially chaotic dynamics. Since it is often impossible to measure the temporal dynamics in qunatum systems, the main and probably the most dramatic manifestation of classical chaos in their phase space is seen in the distribution of spacing between the neighboring energy levels. While the mechanism leading to the onset of chaotic dynamics is different in every system, the level spacing distribution obeys the universal law, changing from Poissonian in the completely integrable systems to Wigner in completely chaotic ones (Fig.\ref{figdist}).
However, despite the fact that the majority of real-world dynamical systems are {\textit{partially chaotic}}, no adequate description of the level statistics in this case have been developed so far. Here we solve this longstanding problem and show that chaos assisted tunneling strongly affects the resulting distribution.}

In the semiclassical limit the eigenmodes of a quantum system closely resemble the trajectories of its classical counterpart. In an integrable system the number of independent time-conserved quantities is equal to the number of degrees of freedom. Each classical trajectory in the phase space lies on a surface with the topological properties of a torus \cite{ReichlBook}. Each torus corresponds to its own set of the values of the integrals of motion. Different modes of an integrable quantum system correspond to different values of quantum numbers and therefore follow different tori; their energies are not correlated, leading to the Poisson distribution of the spacing $s$ between the neighboring energy levels \cite{ReichlBook} (see Fig.\ref{figdist}). 
\begin{equation}
\label{eqPoisson}
P_P(s)=e^{-s}
\end{equation}

With the onset of chaos some tori are destroyed, leading to the formation of the chaotic regions in the phase space. Classical trajectories in these regions diffuse between different (now broken) tori. Correspondingly, different chaotic modes may be represented by a mix of former regular modes, which leads to repulsion between their energies (similarly to repulsion between energies of symmetric and anti-symmetric modes in double-well potential \cite{LandauQM}). In the limit of a ``completely chaotic'' system all modes of the original integrable system are mixed with each other \cite{Berry1977} so that the repulsion exists between any two levels, dramatically changing the distribution from Poissonian to Wigner \cite{ReichlBook} (Fig.\ref{figdist}).
\begin{equation}
\label{eqWigner}
P_W(s)=\frac{\pi}{2}s\, e^{-\frac{\pi}{4}s^2}
\end{equation}

\begin{figure}
\includegraphics[width=4cm]{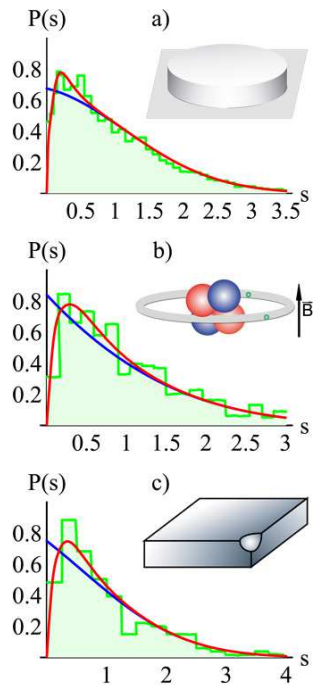}
\caption{\label{figBR}
Level spacing statistics in quantum systems obeys a universal law. We compare the BR formula (blue) and our result, Eq.~(\ref{eqnSpac}) (red) to the data (green) from: 
(a) optical microresonators, where the chaotic ray dynamics is introduced by the deviation of the shape of the cavity from the perfect ellipse; 
(b) Helium atom in a magnetic field, where the dynamics of the electrons becomes partially chaotic since the magnetic field destroys the spherical symmetry of Coulomb potential \cite{SpacingHe};
(c) acoustic resonances in Al blocks, where the dynamics of the acoustic waves becomes chaotic when the shape of the block deviates from parallelepiped \cite{SpacingAl}. Note the suppression of the small-spacings observed in all systems in agreement with the presented theory} 
\end{figure}

While it is straightforward to account for the mode mixing and corresponding level repulsion in a {\it completely} chaotic system using Random Matrix Theory \cite{RMT}, this approach cannot be directly applied to a generic system with mixed regular-chaotic dynamics. The only existing ``first-principle'' approach to this problem neglects the correlation between the regular and chaotic parts of the quantum system leading to the celebrated Berry-Robnik (BR) distribution \cite{BRdistr} 
\begin{eqnarray}
\label{eqnBR}
\nonumber
P_{BR}(s)\propto &\rho^2 e^{-\rho s}\rm{erfc}\left(\frac{\sqrt{\pi}}{2}(1-\rho) s\right)+ \left[
\frac{\pi}{2}(1-\rho)^2 s\right.
\\
& \left.+ 2\rho
\right]
(1-\rho) e^{-\rho s-\frac{\pi}{4}(1-\rho)^2 s^2} 
\end{eqnarray}
where $\rho$ is the relative phase-space volume, occupied by regular trajectories in the mixed system. The limit $\rho\to 1$ corresponds to the regular system, while $\rho\to 0$ represents completely chaotic case. Although BR approach adequately describes the tail of the spacing distribution, it fails to describe the effect of level repulsion at small spacings, present in numerical and experimental data from various partially chaotic physical systems \cite{RobnikRev,SpacingHe,SpacingAl,Stockmann} (Fig.\ref{figBR}). Several, mostly empirical, approaches were suggested to explain this discrepancy \cite{Brody,Hasegawa}, however an adequate description of the physics behind small-splitting behavior of the level statistics still remains an unsolved problem. In the present work we demonstrate that this discrepancy is a manifestation of a special kind of tunneling, known as Chaos-Assisted Tunneling (CAT). 

While the tori represent impenetrable ``dynamical'' barriers for the classical trajectories, the modes of quantum system are allowed to break restrictions imposed by classical mechanics, penetrating ``under'' dynamical barriers. This phenomenon, known as the ``dynamical tunneling''\cite{DT,DTNature,DTScience,CATNar} is common for the multidimensional quantum systems and is similar to tunneling under the potential barrier in 1D systems. As any tunneling process, the dynamical tunneling is exponentially suppressed by the ``effective width'' of the barrier in the phase space. 

\begin{figure}
\includegraphics[width=6cm]{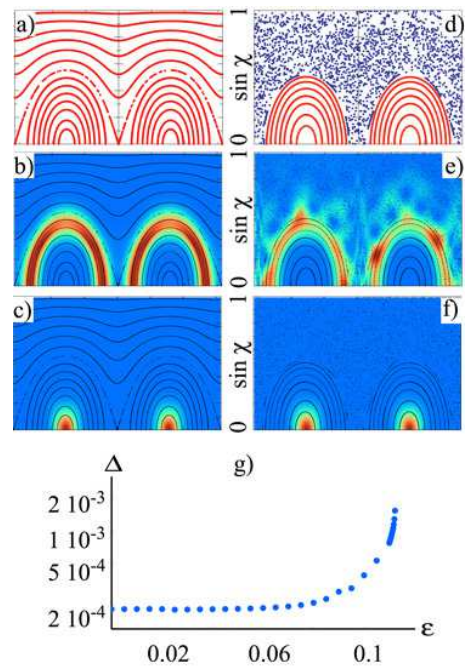}
\caption{\label{figmod}
Ray dynamics is integrable inside an elliptical resonator. Different ray trajectories lay on different tori, which are projected as 1D lines in the Poincar\`{e} surface of section (a). The modes, as visualized by their Husimi projections \cite{ReichlBook}, also closely follow the tori (b,c). As the level repulsion between different modes is absent, the energies of the modes (b) and (c) almost coincide. 
When the shape of the resonator deviates from the perfect ellipse, the ray dynamics inside the cavity becomes partially chaotic. As the deformation increases, so does the amount of chaos in the system. Finally, the tori around mode (c) begin to break down (d). 
Although the torus supporting the mode (b) survives, dynamical tunneling triggers the interaction between the mode (b) and the states which are localized in now chaotic portion of the phase space, leading to the formation of the mode (e). The effect of the transition to partial chaos on the mode (c) is similar, but quantitatively weaker due to a large number of surviving tori surrounding it (f). The result of this indirect ``chaos-assisted'' interaction of the regular modes (b) and (c), is the level repulsion of their successors (e) and (f) - as illustrated in the panel (g) which shows the corresponding energy difference $\Delta$ as a function of deformation (parametrized by $\varepsilon$). 
The strong dependence of $\Delta$ on $\varepsilon$ confirms CAT nature of the level repulsion 
} 
\end{figure}

While the onset of chaotic behavior in the regular system does not cause direct interference of different regular levels, the chaotic levels do interact with the regular ones due to dynamical tunneling, which leads to the repulsion between their energy levels (Fig.\ref{figmod}). Moreover, different regular levels interact (via tunneling) with the same chaotic state, repulsing from each other. This special kind of dynamical tunneling between two regular states via a chaotic one is known as chaos – assisted tunneling \cite{CATNar,CAT}.

\begin{figure}
\includegraphics[width=8cm]{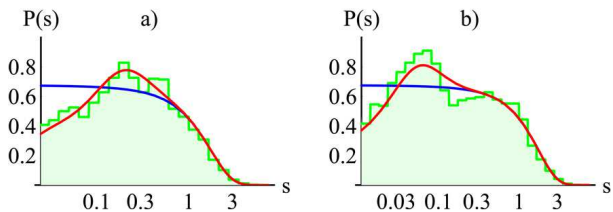}
\caption{\label{figspac}
By changing the ``effective'' Plank constant $\hbar_{\rm eff}$, we change the interaction between quantum states, keeping the classical dynamics unchanged. This leads to a substantial change in small-spacing behavior of the level-spacing distribution [corresponding to change of tunneling parameter $\nu$ in Eq.~(\ref{eqnSpac})], while its large-spacing part is governed by the same classical parameter $\rho$ and is well-described by the BR formula. We compare the level statistics in deformed elliptical resonator (green) to BR formula (blue) and our analytical approach (red); $\hbar_{\rm eff}=1/75$ (a), $\hbar_{\rm eff}=1/100$ (b). (In optical microresonators $\hbar_{\rm eff}=1/kR$, where $k$ is the wavevector inside the cavity and $R$ is average radius of the resonator. The shape of the resonator boundary defines the classical ray dynamics and is kept unchanged)} 
\end{figure}

Because of its tunneling nature, level repulsion leads to a small change in the energy level positions, thus affecting only the small-spacing behavior of the distribution, and leaving large-spacing behavior practically unchanged. Using the perturbative approach similar to that developed in Ref.~[\onlinecite{CAT}], we derive the following expression for the spacing distribution in the quantum system:
\begin{eqnarray}
\nonumber
P(s)\propto&&
\rho^2\,\rm{F}\left(\frac{s}{\nu^2}
\right) e^{-\rho s}
\rm{erfc}\left(\frac{\sqrt{\pi}}{2}(1-\rho) s\right)
+\left[
\frac{\pi
}{2} (1-\rho)^2 s \right.\\
\label{eqnSpac}
&&+\left.2 \rho\, 
\rm{F}\left(
\frac{s}{\nu}
\right)
\right]
(1-\rho)e^{-\rho s-\frac{\pi}{4}(1-\rho)^2 s^2}
\end{eqnarray}
where 
\begin{equation}
\label{eqFunF}
\rm{F}(x)= 1-
\frac{
1-
\sqrt{\frac{\pi}{2}}x}{e^{x}-x}
\end{equation}
In contrast to the parameter $\rho$, corresponding to the {\it classical dynamics} in the system, parameter $\nu$ describes the {\it tunneling between different modes}, so it has intrinsically quantum mechanical nature (see Fig.\ref{figspac}). The effect of the tunneling interaction between different modes of a quantum system on the level spacing distribution can be described by the universal function $\rm{F}(x)$, given by Eq.~(\ref{eqFunF}). The limit $\rm{F}(x)\equiv 1$ which neglects such interaction, leads to BR distribution. Different dependence of function F on the parameter $\nu$ in first and last terms of Eq.(\ref{eqnSpac}) corresponds to {\it direct} and {\it chaos-assisted} tunneling processes respectively. 

In Fig.~\ref{figBR} we illustrate the universality of the developed approach and demonstrate an excellent agreement of our analytical results with experimental and numerical data in a variety of physical systems -- from optical microresonators used in novel microdisk lasers \cite{BTScience} to resonances of atomic Helium in magnetic field \cite{SpacingHe} to acoustic resonances in Al blocks \cite{SpacingAl}. 

\section{Acknowledgements}
Authors acknowledge the support by NSF grant DMR-0134736 and helpful discussions with M.~Berry, S.~Sridhar, A.~D.~Stone, and S.~Tomsovic.

\section{Methods}
{\bf Poincar\'{e} surface of section} (SOS) represents a ``stroboscopic'' projection of the classical trajectories on a 2D-plane. For the case of the optical microresonator we start a ray trajectory at some point inside the cavity and each time the ray hits boundary, we plot the polar angle of the point of incidence and the sine of its angle of incidence. Since the regular trajectories lie on tori, they are constrained to 1D lines in SOS. Chaotic trajectories occupy 2D ``chaotic sea''. 

{\bf Husimi projection} is used to visualize the phase space structure of a wavefunction, projecting it onto coherent state in the phase space \cite{ReichlBook}.


\begin{thebibliography}{}
\bibitem{SpacingHe}K.~Karremans, W.~Vassen, W.~Hogervorst ``Observation of the transition to chaos in the level statistics of diamagnetic helium'', Phys.~Rev.~Lett 81 {\bf 4843} (1998)
\bibitem{DTNature}W.~K.~Heinsenger, H.~H\"{a}ffner, A.~Browaeys, N.~R.~Heckenberg, K.~Helmerson, C.~McKenzie, G.~J.~Milburn, W.~D.~Phillips, S.~L.~Rolson, H.~Ruminsztein-Dunlop, B.~Upcroft ``Dynamical tunneling of ultracold atoms'', Nature, {\bf 412}, 52 (2001) 
\bibitem{DTScience}D.~A.~Steck, W.~H.~Oskay, and M.~G.~Raizen, ``Observation of chaos-assisted tunneling between islands of stability'', Science {\bf 293}, 274 (2001) 
\bibitem{BTScience} C.~Gmachl, F.~Capasso, E.~E.~Narimanov, J.~U.~N\"{o}ckel, A.~D.~Stone, J.~Faist, D.~L.~Sivco, A.~Y.~Cho, ``High-power directional emission from microlasers with chaotic resonators'', Science {\bf 280}, 1556 (1998)
\bibitem{Wells}T.~M.~Fromhold, L.~Eaves, F.~W.~Sheard, M.~L.~Leadbeater, T.~J.~Foster, P.~C.~Main ``Mabetotunneling spectroscopy of a quantum well in the regime of classical chaos'', Phys.~Rev.~Lett.~ {\bf 72} 2608 (1994); G.~M\"{u}ller, G.~S.~Boebinger, H.~Mathur, L.~N.~Pfeiffer, and K.~W.~West, ``Precursors and transition to chaos in quantum well in a tilted magnetic field'', Phys.~Rev.~Lett {\bf 75} 2875 (1995); E.~E.~Narimanov, A.~D.~Stone, G.~S.~Boebinger, ``Semiclassical theory of magnetotransport through a chaotic quantum well'', Phys.~Rev.~Lett. {\bf 80} 4024 (1998); 
\bibitem{Dots}R.~A.~Jalabert, A.~D.~Stone,Y.~Alhassid, ``Statistical theory of Coulomb blocade oscillations: Quantum chaos in quantum dots'', Phys.~Rev.~Lett.~ {\bf 68} 3468 (1992); A.~M.~Chang, H.~U.~Baranger, L.~N.~Pfeiffer, K.~W.~West, T.~Y.~Chang, ``Non-Gaussian distribution of coulomb blockade peak heights in quantum dots'', Phys.~Rev.~Lett {\bf 76} 1695 (1995); E.~E.~Narimanov, N.~R.~Cerruti, H.~U.~Baranger, S.~Tomsovic, ``Chaos in Qunatum Dots: Dynamical Modulation of Coulomb blocade peak heights'', Phys.~Rev.~Lett, {\bf 83} 2640 (1999); S.~M.~Cronecwett, S.~R.~Patel, C.~M.~Marcus, K.~Campman, A.~C.~Gossard, ``Mesoscopic fluctuations of elastic cotunneling in coulomb blockaded quantum dots'', Phys.~Rev.~Lett {\bf 79} 2312 (1997) 
\bibitem{WGNature}J.~U.~N\"{o}ckel, A.~D.~Stone, ``Ray and wave chaos in asymmetric resonant optical cavities'', Nature {\bf 385}, 45 (1997)
\bibitem{KodroliSridhar1994}A.~Kudroli,S.~Sridhar, A.~Pandey, R.~Ramaswamy ``Signatures of chaos in quantum billiards: Microwave experiments'', Phys.~Rev.~E {\bf R11} 49 (1994); A.~Kudrolli, V.~Kidambi, and S.~Sridhar, ``Experimental studies of chaos and localization in quantum wavefunctions'', Phys.~Rev.~Lett {\bf 75} 822 (1995)
\bibitem{Alt1994}H.~Alt, H.~-D.~Gr\"{a}ft, H.~L.~Harney, R.~Hofferbert, H.~Lengeler, C.~Rangacharyulu, A.~Richter, P.~Schardt, ``Superconducting billiard cavities with chaotic dynamics: An experimental test of statistical measures'', Phys.~Rev.~E {\bf R1} 50 (1994)
\bibitem{SpacingAl}C.~Ellegaard, T.~Guhr, K.~Lindemann, H.~Q.~Lorensen, J.~Nyg\.~{a}rd, M.~Oxborrow, ``Spectral statistics of acoustic resonances in aluminum blocks'', Phys.~Rev.~Lett.~ {\bf 1546} 75 (1995)
\bibitem{ReichlBook} L.~E.~Reichl {\it The transition to chaos in conservative classical systems: Quantum manifestations}, Springer-Verlag (1992)
\bibitem{Stockmann}H.~J.~St\"{o}ckmann, {\it``Quantum Chaos: an Introduction''}, Cambridge University Press, Cambridge 1999; H.~J.~St\"{o}ckmann, J.~Stein, ``Quantum chaos in billiards studied by microwave absorption'', Phys.~Rev.~Lett. {\bf 64} 2215 (1990)
\bibitem{Berry1977}M.~Berry ``Regular and irregular semiclassical wavefunctions'', J.~Phys.~A {\bf 10} 2083 (1977)
\bibitem{LandauQM}L.~D.~Landau, E.~M.~Lifshitz {\it Quantum Mechanics (Non-relativistic theory)},Butterworth Heinemann (1981)
\bibitem{RMT}M.~L.~Mehta {\it ``Random Matrices and statistical theory of energy levels''}, Academic, NY 1967
\bibitem{BRdistr}M.~V.~Berry, M.~Robnik, ``Semiclassical level spacing when regular and chaotic orbits coexist'', J.~of Physics {\bf A17} 2413 (1984)
\bibitem{RobnikRev}H.~Hasegawa, M.~Robnik, G.~Wunner {\it Classical and Quantum Chaos in the Diamagnetic Kepler Problem} Progress of Theoretical Physics Suppl.~ {\bf 98} (1989)
\bibitem{Brody}T.~A.~Brody ``A statistical measure for repulsion of energy levels'', Lett.~Nuovo Cim.~ {\bf 7}, 482 (1973)
\bibitem{Hasegawa}H.~Hasegawa, H.~J.~Mikeska, H.~Frahm, ``Stochastic formulation of energy-level statistics'', Phys.~Rev.~A {\bf 38} 395 (1988)
\bibitem{CATNar}H.~E.~Tureci, H.~G.~L.~Schwefel, A.~D.~Stone, E.~E.~Narimanov, 
``Gaussian-optical approach to stable periodic orbit resonances of partially chaotic dielectric microcavities'', Optics Express {\bf 10}, 752 (2002)
\bibitem{CAT}V.~A.~Podolskiy, and E.~E.~Narimanov ``Semiclassical Description of Chaos-Assisted Tunneling'', accepted to Phys.~Rev.~Lett
\bibitem{DT}M.~J.~Davis and E.~J.~Heller, ``Quantum dynamical tunneling in bound states'', J.~Chem.~Phys. {\bf75}, 246 (1981)
\end{thebibliography}
\end{document}